\documentclass[conference]{IEEEtran}
\IEEEoverridecommandlockouts
\usepackage{amsmath,amssymb,amsfonts}
\usepackage{listings}
\usepackage{algorithm}
\usepackage{algorithmicx}
\usepackage[noend]{algpseudocode}
\usepackage{graphicx}
\usepackage{textcomp}
\usepackage{xcolor}
\usepackage{tikz}
\usetikzlibrary{shapes,positioning,calc,arrows,chains}
\usepackage[unicode, hidelinks]{hyperref}
\usepackage[hyphenbreaks]{breakurl}
\usepackage[style=ieee, bibencoding=utf8, citestyle=numeric-comp]{biblatex}
\def\BibTeX{{\rm B\kern-.05em{\sc i\kern-.025em b}\kern-.08em
    T\kern-.1667em\lower.7ex\hbox{E}\kern-.125emX}}
\usepackage{booktabs}
\usepackage{colortbl}
\usepackage{multirow}
\usepackage{flushend}
\usepackage{verbatim}

\algnewcommand{\algorithmicand}{\textbf{ and }}
\algnewcommand{\algorithmicor}{\textbf{ or }}
\algnewcommand{\OR}{\algorithmicor}
\algnewcommand{\AND}{\algorithmicand}

\definecolor{light-gray}{gray}{0.80}

\begin{document}

\title{Strong Optimistic Solving for \\
Dynamic Symbolic Execution
\thanks{This work was supported by RFBR grant 20-07-00921 A.}}

\author{
\IEEEauthorblockN{
  Darya Parygina\IEEEauthorrefmark{1}\IEEEauthorrefmark{2},
  Alexey Vishnyakov\IEEEauthorrefmark{1} and
  Andrey Fedotov\IEEEauthorrefmark{1}
}
\IEEEauthorblockA{
  \IEEEauthorrefmark{1}Ivannikov Institute for System Programming of the RAS
}
\IEEEauthorblockA{
  \IEEEauthorrefmark{2}Lomonosov Moscow State University
}
Moscow, Russia \\
\{pa\_darochek, vishnya, fedotoff\}@ispras.ru
}

\maketitle

\begin{tikzpicture}[remember picture, overlay]
\node at ($(current page.south) + (0,0.65in)$) {
\begin{minipage}{\textwidth} \footnotesize
 Parygina D., Vishnyakov A., Fedotov A. Strong Optimistic Solving for Dynamic
 Symbolic Execution. 2022 Ivannikov Memorial Workshop (IVMEM), IEEE, 2022, pp.
 43-53. DOI: \href{https://www.doi.org/10.1109/IVMEM57067.2022.9983965}{10.1109/IVMEM57067.2022.9983965}.

 \copyright~2022 IEEE. Personal use of this material is permitted. Permission
 from IEEE must be obtained for all other uses, in any current or future media,
 including reprinting/republishing this material for advertising or promotional
 purposes, creating new collective works, for resale or redistribution to
 servers or lists, or reuse of any copyrighted component of this work in other
 works.
\end{minipage}
};
\end{tikzpicture}

\begin{abstract}
Dynamic symbolic execution (DSE) is an effective method for automated program
testing and bug detection. It is increasing the code coverage by the complex
branches exploration during hybrid fuzzing. DSE tools invert the branches along
some execution path and help fuzzer examine previously unavailable program
parts. DSE often faces over- and underconstraint problems. The first one leads
to significant analysis complication while the second one causes inaccurate
symbolic execution.

We propose strong optimistic solving method that eliminates irrelevant path
predicate constraints for target branch inversion. We eliminate such symbolic
constraints that the target branch is not control dependent on. Moreover, we
separately handle symbolic branches that have nested control transfer
instructions that pass control beyond the parent branch scope, e.g. return,
goto, break, etc. We implement the proposed method in our dynamic symbolic
execution tool Sydr.

We evaluate the strong optimistic strategy, the optimistic strategy that
contains only the last constraint negation, and their combination. The results
show that the strategies combination helps increase either the code coverage or
the average number of correctly inverted branches per one minute. It is optimal
to apply both strategies together in contrast with other configurations.
\end{abstract}


\section{Introduction}

Modern software developers invest a huge amount of efforts in increasing the programs quality. Companies employ security development lifecycle (SDL)~\cite{howard06, iso08, gost16} to find bugs in their products and defend programs from dangerous interventions. Various automated testing tools provide thorough code exploration. Hybrid fuzzing~\cite{pak12} is the state-of-the-art solution for finding bugs. Its efficiency comes from fast and lightweight fuzzing~\cite{serebryany16, fioraldi20}, that allows to discover new paths quickly, and more accurate dynamic symbolic execution~\cite{stephens16, yun18, vishnyakov20, poeplau20, poeplau21, borzacchiello21}, that helps reaching difficult program parts by inverting complicated branches along some execution path. Moreover, dynamic symbolic execution (on initially valid paths) enables new seeds generation that trigger memory and undefined behavior errors~\cite{vishnyakov21}.

Dynamic symbolic execution~\cite{king76, baldoni18} tools execute the target program and construct the path predicate by symbolic interpretation of program instructions. The path predicate contains the branch conditions met during analysis. Symbolic engines try to invert each branch from the path predicate to discover new execution paths that are hardly reached with fuzzing. The predicate for the branch inversion conjuncts all the preceding branch constraints (i.e. constraints from branches executed before the target branch) and the negation of the target branch constraint. The majority of symbolic engines frequently face over- and underconstraint problems (similar to taint analysis~\cite{kang11}) that prevent them from exploring more program paths. Overconstraint denotes the situation when there are many redundant constraints in the path predicate that may cause its complication or even unsatisfiability. Overconstraint is increasing the symbolized instructions number during analysis. Underconstraint, on the contrary, means that some variable is not treated as symbolic, though it should be. Non-trivial branch conditions or symbolic pointers (that depend on user data) may cause underconstraint.

Symbolic executors apply various techniques to cope with these problems (Section~\ref{related-work}). We propose the strong optimistic solving method that generates input data for branch inversion even when the constructed predicate is unsatisfiable. We eliminate some constraints from the solver query based on the branches nesting. Thus, we can bypass over- and underconstraint.

As an illustration, consider the code example below:

\begin{lstlisting}[language=C, basicstyle=\small\ttfamily, numbers=left, xleftmargin=2em, caption={Strong optimistic solving.}, captionpos=b, label=lst:example]
void func(const char* buf) {
  if (buf[3] == '6' & buf[0] == '5') //c_2
    printf("Success!\n");
  else
    printf("Fail\n");
}

int main() {
  char buf[4];
  read(0, buf, 4);
  if (buf[2] < '0')                  //c_11
    printf("Eliminated by slicing\n");
  if (buf[0] == '3')                 //c_13
    printf("Independent branch\n");
  if (buf[1] - buf[3] == 1)          //c_15
    func(buf);
}
\end{lstlisting}

Suppose the execution got to line~5 and the branch conditions in lines~11, 13, and 15 appeared to be true, while the branch constraint in line~2 turned out to be false. Our goal is inverting the branch in \texttt{func} and reaching line~3. As we want if-statement in line~2 to be true, the corresponding predicate will have the following form:
$(\texttt{buf[2]} < \texttt{'0'})\ \land\ (\texttt{buf[0]} = \texttt{'3'})\ \land\ (\texttt{buf[1]} - \texttt{buf[3]} = 1)\ \land\ (\texttt{buf[3]} = \texttt{'6'}\ \land\ \texttt{buf[0]} = \texttt{'5'})$.

This predicate is unsatisfiable, since \texttt{buf[0]} cannot be equal to \texttt{'3'} and \texttt{'5'} simultaneously. The path predicate is overconstrained, as the branches in lines~11 and 13 do not actually affect inversion of the branch in line~2. Many real-world programs face similar situations, and the number of irrelevant constraints in the path predicates grows along with the code expansion.

This paper makes the following contributions:
\begin{itemize}
  \item We propose strong optimistic solving that allows to discover new program paths via dynamic symbolic execution. We propose a new strategy for eliminating irrelevant symbolic constraints from the path predicate, based on program call stack and branch control dependency analysis. We propose an improvement to this strategy that avoids eliminating constraints which may potentially affect the program control flow. We leave the corresponding symbolic branches in the path predicate, thus increasing the chance that the generated test case will actually invert the target branch.
  \item We evaluate the method on the set of real-world applications~\cite{sydr-benchmark}, measuring the analysis efficiency (i.e. correctly inverted branches number, accuracy, and speed) and the code coverage. We also evaluate optimistic solving strategy~\cite{yun18} and the methods combination. We compare the results for different configurations. We show that the methods combination helps to either increase the code coverage by inverting more symbolic branches along the single execution path, or improve symbolic execution efficiency.
\end{itemize}

The rest of this paper is organized as follows. Section~\ref{related-work} contains the overview of techniques, applied in symbolic executors to cope with over- and underconstraint problems. Section~\ref{str-opt-description} presents the strong optimistic solving strategy. Implementation details are discussed in section~\ref{implementation}. The experimental evaluation is shown in
Section~\ref{evaluation}. Finally, Section~\ref{conclusion} concludes this paper.

\section{Related Work}
\label{related-work}

Overconstraint problem leads to significant difficulties during symbolic execution. Excessive constraints complicate the path predicate and, in some cases, make it unsatisfiable. Underconstraint is another important obstacle while constructing the path predicate. The common reason for underconstraint is related to the symbolic addresses problem, i.e. when load or store address depends on user input. In this section we systematically analyze the variety of methods, used in symbolic engines to cope with over- and underconstraint.

\subsection{Irrelevant Constraint Elimination}

Symbolic variables and symbolic branches abundance frequently results in great number of redundant constraints in the path predicate. Irrelevant constraint elimination methods intend to remove excessive constraints from the solver queries.

In KLEE~\cite{cadar06, cadar08} the constraint independence optimization exploits the fact that the set of constraints can be often divided into several subsets of independent constraints. Two constraints are independent when they are affected by disjoint sets of symbolic variables. Building independent constraints subsets is performed by constructing the special graph. Each pair of connected graph nodes illustrates symbolic variables that are included in the same symbolic constraint. Each independent constraints subset is built by adding constraints, containing symbolic variables from the certain connected component of the graph. Such optimization benefits from reducing solver load by discarding irrelevant constraints when querying for the certain constraint satisfiability. Even in the worst case, when no constraints can be eliminated, the query cost is the same as without optimization applied (the cost of computing the independent subsets is small and can be omitted).

The path predicate slicing~\cite{vishnyakov20}, proposed in our previous work and implemented in Sydr, is also used for reducing solver load when inverting the target branch. Sydr eliminates irrelevant path predicate conjuncts from the queries. We consider some constraint irrelevant if its symbolic variables do not directly or transitively depend on symbolic variables from the target branch constraint. Symbolic variable $var_1$ directly depends on variable $var_2$ if there is some constraint in the path predicate that contains these variables simultaneously. Similarly, variable $var_1$ transitively depends on variable $var_2$ if $var_1$ depends directly on some other variable that, in its turn, depends directly or transitively on $var_2$.

At the first step, Sydr determines the set of symbolic variables that directly or transitively depend on the target branch constraint variables. This set contains all the symbolic data that are responsible for the target branch inversion. After that, the new sliced predicate is formed from only those constraints that contain the selected variables. Thus, solver-generated model will contain only a subset of all symbolic data. The values for the missing variables are retrieved from the initial input data as they are already the solution for the path predicate. Applying slicing leads to a more powerful symbolic execution with less time and memory consumption during queries solving.

\subsection{Optimistic Solving}

The hybrid fuzzer QSYM~\cite{yun18} suggests the heuristic method called optimistic solving. If the whole path predicate is not satisfiable, QSYM selects just the last constraint and tries to get its solution. The last constraint often has a simple form that facilitates its solving. Additionally, test cases, generated by the optimistic solving, have a good chance to actually invert the target branch, since they satisfy at least the last (target branch) constraint. This heuristic allows QSYM to explore new code quite efficiently due to the fact that the emulation overhead dominate the overhead for constraint solving in the real-world applications. Optimistic solving provides much more correct test cases for inverting symbolic branches at the same time.

\subsection{Function Semantics Modeling}

During native symbolic execution, all instructions are instrumented and all symbolic branch constraints are added to the path predicate. Nevertheless, such strategy often leads to the great time costs for symbolic execution of instrumented code and significant path predicate expansion, especially while processing library calls. In order to manage these problems, symbolic engines apply various techniques allowing to decrease the number of redundant constraints in the path predicate and reduce the amount of symbolic instructions.

The library code is an essential part of real-world applications. But at the same time, it is a reason for the great increase in the symbolic instructions number. The function arguments symbolization makes the statements inside the function implementation also symbolized if they are affected by these arguments. The key intuition of the function semantics modeling methods is that there is no need to symbolically execute every internal instruction in library functions with standard specified semantics. Such approach benefits not only from speeding up symbolic execution, but also from diminishing symbolic state overconstrainting.
There are five main directions in function semantics modeling.

The first one, implemented in KLEE~\cite{cadar08}, proposes to replace libc functions with their simpler precompiled LLVM IR versions. Such approach simplifies the code and avoids extra components compilation. However, it does not consider complete function semantics.

The second method, applied in SymCC~\cite{poeplau20}, adds additional constraints to the path predicate instead of symbolically executing the function code. These constraints model the concrete execution trace.

Another approach is modeling each function with a single branch that represents different function results. For instance, \texttt{strcmp} strings may be equal or not. Such method is one of the three function semantics modeling strategies implemented in FUZZOLIC~\cite{borzacchiello21}.

The fourth strategy focuses on construction of symbolically computed expression for the return value. Under this approach, in S$^2$E~\cite{chipounov11}, the special wrappers insert the symbolic formulas in the return values. The similar method is applied in Sydr~\cite{vishnyakov20} where the construction of symbolic formulas is performed for the set of complicated functions that deal with comparisons and string to integer
conversions; and in angr~\cite{shoshitaishvili16} where some functions are replaced by the simplified implementations while others are fully modeled according to their semantics.

The last but not least method suggests simply skipping functions that often contain symbolized arguments but do not affect the branches inversion, e.g. heap operations (\texttt{malloc}, \texttt{calloc}, \texttt{realloc}, \texttt{free}, etc.), printing functions like \texttt{printf}. Such functions append lots of inconsequential constraints, thus can be skipped without the soundness loss.

\subsection{Index-Based Memory Model}

There are two common approaches for handling symbolic pointers: concretizing symbolic indexes and using a fully symbolic memory model. The first one leads to the underconstraint problem while the second one causes overconstraint and high performance costs.

Mayhem~\cite{cha12} proposes an index-based memory model as an approach to handling symbolic addresses according to the memory index value. Due to the performance issues, Mayhem uses not the full symbolic memory model, but the partial one where write addresses are always concretized and reads are treated as symbolic. The similar approach is implemented in Sydr~\cite{kuts21}.

For symbolic reads modeling, Mayhem introduces memory objects that map indexes to symbolic expressions. Each memory object bounds are determined via solver with a binary search. Mayhem applies some optimizations to lighten the burden on the solver during this process. Value set analysis~\cite{balakrishnan04} allows to reduce the set of appropriate values in the memory object index interval by replacing it with a strided interval. Each strided interval is then refined using solver queries.
Moreover, Mayhem uses balanced index search trees (IST), i.e. binary search trees with the index as a key and the memory object entries as leaf nodes.
As an effective approach to reducing the IST nodes number, Mayhem applies bucketization with linear functions which key idea is combining multiple memory object entries into a single index-parametrized leaf: $k * i + b$.


\subsection{Basic Block Pruning}

The problem of adding constraints, repetitively generated by the same code, to the path predicate is also crucial for symbolic execution. It is especially relevant when processing loops and function calls. Such constraints prevent new code exploration and can even make solver queries unsatisfiable due to their computational complexity. QSYM~\cite{yun18} applies the method called basic block pruning to cope with this issue. Its essence is detecting the repetitive basic blocks and pruning them from symbolic execution. The execution frequency of each basic block is measured at runtime (each time the block is executed, its frequency counter is incremented), and QSYM stops generating further constraints from basic blocks that have been executed too frequently. Exponential back-off is used to prune rapid blocks (the blocks whose frequency number is a power of 2 are the only executed blocks) while heuristic approaches are applied to avoid excessive pruning. The first heuristic is grouping multiple executions. Thus, several executions make a single increment to the frequency counter. And the second one is context-sensitivity that distinguishes frequency counters for the same basic block in different call contexts.

\subsection{Taint-Assisted Partial Symbolization}

As a way to cope with the overconstraint problem, LeanSym~\cite{xianya21} proposes taint-assisted partial symbolization. This approach has similar purposes as the path predicate slicing, yet exploits the fact that concolic execution involves a concrete input and uses it to compute the data flow.

LeanSym applies dynamic taint analysis (DTA) to identify input bytes affecting different symbolic branches. At the beginning, all the input data are tainted and the direct taint propagation is performed. As a result, input bytes affecting each branch \texttt{cmp} operands are determined. On the next analysis stage, LeanSym begins inverting symbolic branches along the execution path. It selects the target branch and symbolizes only those bytes that affect its \texttt{cmp} operands. These bytes offsets are taken from the previous DTA run results. Finally, LeanSym symbolically executes only the statements that operate on the relevant symbolic data. The path predicate is updated with the constraints that are necessary for the target branch inversion. Values for the symbolized input bytes are taken from the solver-generated model while the rest values are retrieved from the initial input.

Taint-assisted partial symbolization provides a good speed increase of symbolic execution and queries solving. But at the same time, it may lead to enlarging the number of unsatisfiable queries due to non-symbolized bytes concretization, leading to the underconstraint.

\section{Strong Optimistic Strategy}
\label{str-opt-description}

New paths exploration is one of the main symbolic execution purposes. A good strategy for increasing the code coverage is inverting symbolic branches along some execution trace. But due to the problems of over- and underconstraint, the solver queries for inverting symbolic branches in a large number of cases turn out to be unsatisfiable. The vast majority of symbolic executors make no attempts to reuse already collected symbolic constraints for inverting the target branch or use them in a very simple form. For instance, optimistic solutions in QSYM~\cite{yun18} do not tend to satisfy symbolic constraints needed for reaching the target point in many real-world applications.

Path predicate slicing, implemented in Sydr~\cite{vishnyakov20}, is a quite effective method of reusing collected symbolic constraints to explore new code. It proposes partial constraint selection based on data dependencies on the target branch. But despite an intention to minimize the path predicate and save its soundness, slicing still can provide unsatisfiable predicates. However, some constraints can still be removed without accuracy loss. Hence, as a continuation of the path predicate slicing, we introduce the strong optimistic strategy that allows to explore new code by removing excessive symbolic constraints from the sliced path predicate in attempt to invert the target branch, thus exploring previously non-invertible symbolic branches.

In our work we base on binary code analysis for x86 architecture. Thus, we handle programs written in compiled languages such as C, C++, Rust, Go, etc. We analyze the unmodified binary code and ignore the program source code. The proposed method is independent of most compiler optimizations, except for gcc with \texttt{O1} optimization level, as in this case the \texttt{if}-statement is compiled in unusual way. For example, consider the binary code generated with gcc \texttt{O0/O2/O3} (Listing~\ref{lst:if}) and \texttt{O1} (Listing~\ref{lst:if-o1}) for the fragment \texttt{if~(x~>~0) printf(...);}. In Listing~\ref{lst:if} \texttt{printf} call goes right after the conditional branch. However, in Listing~\ref{lst:if-o1} the \texttt{printf} call is located in separate basic block that jumps back to the instruction following the conditional branch.

\begin{lstlisting}[basicstyle=\scriptsize\ttfamily, xleftmargin=2em, escapechar=!, label=lst:if, caption={gcc -O0/O2/O3 code.}, captionpos=b]
 724:   mov    eax,DWORD PTR [rbp-0xc]
 727:   test   eax,eax
 729:   jle    737
 72b:   lea    rdi,[rip+0xc5]
 732:   call   5d0 <puts@plt>
 737:   ...
\end{lstlisting}

\begin{lstlisting}[basicstyle=\scriptsize\ttfamily, xleftmargin=2em, escapechar=!, label=lst:if-o1, caption={gcc -O1 code.}, captionpos=b]
 71a:   cmp    DWORD PTR [rsp+0x4],0x0
 71f:   jg     73b
 721:   ...
 ...
 73a:   ret
 73b:   lea    rdi,[rip+0xb5]
 742:   call   5d0 <puts@plt>
 747:   jmp    721
\end{lstlisting}

The key intuition of the strong optimistic strategy is that symbolic branches, which the target branch is not control dependent on, do not affect this branch inversion, as they can be executed independently of it. Each program branch has source ($src\_addr$) and destination ($dst\_addr$) addresses, where the first one is the corresponding \texttt{jcc}-instruction address and the second one is the jump destination address. We consider that the first branch is control dependent on the second (or the first branch is nested in the second) if $src\_addr_2 < src\_addr_1 < dst\_addr_2$. For example, if we have a symbolic branch, such that its destination address is less than the target branch source address, this branch is independent of the target one. Furthermore, when there is a function call before the target branch and there are symbolic branches inside this function, such branches have no impact on the target branch inversion. So, if we remove symbolic branches, that the target branch is not control dependent on, from the sliced predicate, we may have a better chance of getting a satisfiable predicate.

Consider the example in Listing~\ref{lst:example}. The whole path predicate, that contains symbolic constraints from lines~11, 13, 15, and 2, is unsatisfiable, as described above. Applying path predicate slicing is not enough in this case: constraints in lines~13, 15, and 2 depend directly on symbolic variables from the target constraint (\texttt{buf[3]} and \texttt{buf[0]}). So, the sliced predicate still contains conflicting symbolic constraints from lines~2 and 13. On the other hand, the optimistic strategy is not enough too, since the constraint in line~15 can occur to be false and the target branch will not be reached. But if we notice that the first constraint in the sliced path predicate (in line~13) corresponds to the independent branch and remove it from the predicate, it becomes satisfiable. The destination address of branch in line~13 is line~15. Thus, neither branch in line~15, nor branch in line~2 is control dependent on branch in line~13, so, branch in line~13 can be eliminated without any consequences for the target branch inversion. Hence, the only correct solution for the symbolic data can be found with the strong optimistic strategy: $\texttt{buf[0]} = \texttt{'5'},\ \texttt{buf[1]} = \texttt{'7'},\ \texttt{buf[3]} = \texttt{'6'}$.

\begin{figure}[htbp]
\centering
\scalebox{0.5}
{
\colorlet{lcfree}{black}
\colorlet{lcnorm}{black}
\colorlet{lccong}{black}

\pgfdeclarelayer{marx}
\pgfsetlayers{main,marx}

\providecommand{\cmark}[2][]{%
  \begin{pgfonlayer}{marx}
    \node [nmark] at (c#2#1) {#2};
  \end{pgfonlayer}{marx}
  }
\providecommand{\cmark}[2][]{\relax}
\begin{tikzpicture}[%
    >=triangle 60,              
    start chain=going below,    
    node distance=6mm and 60mm, 
    every join/.style={norm},   
    ]

\tikzset{
  base/.style={draw, on chain, on grid, align=center, minimum height=4ex},
  proc/.style={base, rectangle, text width=8em},
  test/.style={base, diamond, aspect=2, text width=5em},
  term/.style={proc, rounded corners},
  coord/.style={coordinate, on chain, on grid, node distance=6mm and 25mm},
  nmark/.style={draw, cyan, circle, font={\sffamily\bfseries}},
  norm/.style={->, draw, lcnorm},
  free/.style={->, draw, lcfree},
  cong/.style={->, draw, lccong},
  it/.style={font={\small\itshape}}
}

\node [test] (t1) {Sliced predicate SAT?};
\node [test, join] (t2) {Optimistic predicate SAT?};

\node [term, fill=lcfree!25, right=of t1] (p1) {Save test case};
\node [test, right=of t2] (t3) {Strong optimistic matches optimistic?};
\node [test] (t4) {Strong optimistic predicate SAT?};
\node [term, fill=lcfree!25, join] (p2) {Save optimistic};

\node [term, fill=lcfree!25, right=of t3] (p3) {Save optimistic};
\node [term, fill=lcfree!25, right=of t4] (p4) {Save optimistic, strong\_optimistic};

\node [coord, right=of t1] (c1)  {};
\node [coord, right=of t2] (c2)  {};
\node [coord, right=of t3] (c3)  {};
\node [coord, right=of t4] (c4)  {};

\path (t1.south) to node [near start, xshift=1em, yshift=-0.5em] {$no$} (t2);
  \draw [->,lcnorm] (t1.south) -- (t2);
\path (t3.south) to node [near start, xshift=1em, yshift=-0.5em] {$no$} (t4);
  \draw [->,lcnorm] (t3.south) -- (t4);
\path (t4.south) to node [near start, xshift=1em, yshift=-0.5em] {$no$} (p2);
  \draw [->,lcnorm] (t4.south) -- (p2);

\path (t1.east) to node [above, xshift=2.5em] {$yes$} (c1);
  \draw [->,lccong] (t1.east) -- (p1);
\path (t2.east) to node [above, xshift=0.8em] {$yes$} (c2);
  \draw [->,lccong] (t2.east) -- (t3);
\path (t3.east) to node [above, xshift=2.5em] {$yes$} (c3);
  \draw [->,lccong] (t3.east) -- (p3);
\path (t4.east) to node [above, xshift=2.5em] {$yes$} (c4);
  \draw [->,lccong] (t4.east) -- (p4);

\end{tikzpicture}
}
\caption{Strong optimistic strategy flowchart.}
\label{flowchart}
\end{figure}
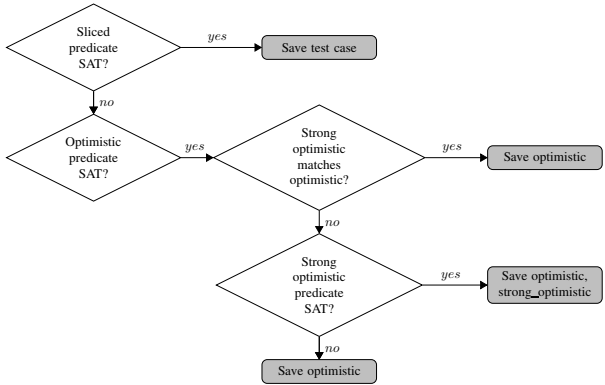

We propose the strong optimistic solving for more accurate symbolic branches inversion and the code coverage increase. The method flowchart is shown in Figure~\ref{flowchart}. If the original path predicate for target branch inversion is satisfiable, then no strategy is applied in order to avoid resources waste. Otherwise, we try to solve only the last constraint, i.e. use the optimistic strategy. If it appears to be SAT, we apply the strong optimistic strategy and try to get a strong optimistic solution. If the optimistic predicate is unsatisfiable, it means that it makes no sense to even try applying another strategies, because the target constraint will always make the modified predicate unsatisfiable.

According to the flowchart in Figure~\ref{flowchart}, we should construct the strong optimistic predicate only if the sliced predicate is UNSAT and the optimistic predicate is SAT. So, the goal is to get something in between these variants, such that it is more likely to be SAT unlike the first one and more likely to be correct unlike the second one. The key intuition is that we should eliminate such symbolic branches that the target branch is not control dependent on. In other words, if some branch is related to the \texttt{src\_addr} $\rightarrow$ \texttt{dst\_addr} jump, then the fact, that the target branch is control dependent on it, is equivalent to the condition \texttt{src\_addr $\leq$ target\_branch\_addr < dst\_addr}. It is worth noting that sometimes strong optimistic predicate may be the same as optimistic predicate, then we should not query solver with the same predicate twice.

The Algorithm~\ref{alg:str-opt} shows how the strong optimistic predicate is constructed. We analyze the program call stack to determine the symbolic branches from the sliced path predicate that the target branch is control dependent on. The global \texttt{call\_stack} structure, passed as input to the algorithm, contains the mapping from each branch constraint in the original predicate $\Pi$ to the program call stack at this branch source address. The program call stack in its turn holds call sites for the functions called before the certain program point.

\begin{algorithm}[htbp]
\small
  \caption{Strong optimistic predicate construction.}
  \textbf{Input:} $call\_stack$~-- program call stack,
  $point$~-- target branch address, $c_{br}$~-- target branch constraint, $\Pi$~-- sliced path predicate for inverting target branch.

  \textbf{Output:} $\Pi_{sopt}$~-- strong optimistic predicate for inverting target branch.

  \begin{algorithmic}
    \State $\Pi_{sopt} \gets c_{br}$
    \Comment{strong optimistic predicate}
    \State $cs \gets call\_stack(point)$
    \ForAll {$c \in reversed(\Pi)$}
        \If {$call\_stack(c)$ is not a prefix of $cs$}
            \State \textbf{continue}
        \EndIf
        \If {$call\_stack(c)$ is prefix of $ cs$ \AND $call\_stack(c) \neq cs$}
            \State $point \gets cs[size(call\_stack(c))]$
            \State $cs \gets call\_stack(c)$
        \EndIf
        \If {$src(c) \leq point < dst(c)$ \OR \texttt{ret} / \texttt{far jmp} / \texttt{far jcc} inside $(src(c),~dst(c))$}
            \State $\Pi_{sopt} \gets \Pi_{sopt} \land c$
        \EndIf
    \EndFor
    \State \textbf{return} $\Pi_{sopt}$
  \end{algorithmic}
  \label{alg:str-opt}
\end{algorithm}

Before the beginning of the algorithm, we choose the target branch with the target constraint $c_{br}$ and apply the path predicate slicing. At the first step, we initialize the resulting predicate $\Pi_{sopt}$ with the target constraint $c_{br}$ and save the target branch call stack in $cs$. Then we start iterating over the sliced predicate constraints in the reversed order. This way, we move from near to far constraints in the direction of the program call stack reduction. At each iteration only three variants are possible: call stack for the analyzed constraint is not a prefix of the current call stack $cs$, or it is a strict prefix, or they are equal. The first option means that the analyzed branch is located in the function that was called and returned before the target branch, thus the target branch is not control dependent on it. The left two variants may potentially cause the target branch to be control dependent on the analyzed branch, and we have to make additional checks in that cases. If we get the second variant, before any other actions we should change the target $point$ and the current call stack $cs$ as constraint $c$ is located in different function frame compared to the current $point$. For this reason, we reassign $point$ with the call site of the first function distinguishing $call\_stack(c)$ and $cs$, and save the value of $call\_stack(c)$ in $cs$. It means that we will check  whether this function call site is control dependent on the following analyzed branches in order to guarantee that the function is called.

When the $point$ and analyzed constraint are located in the same function frame, we start making additional checks. The main reason for the constraint to be added to the strong optimistic predicate is the fact that the point is nested in the corresponding branch. This check is denoted as $src(c) \leq point < dst(c)$, where $src$ and $dst$ return the respective jump source and destination addresses. But there is another important condition notifying that constraint should not be eliminated: the corresponding branch scope contains at least one control transfer instruction (CTI). There are two main instruction types in the CTI category. Firstly, all variants of \texttt{ret} instruction as they lead to premature function termination. Secondly, conditional and unconditional jumps with the destination addresses beyond the analyzed branch scope borders. We process such instructions only if their destination address is greater than the last instruction address in the scope. On the contrary, we do not process backward jumps because it leads to worse symbolic execution accuracy and speed. The two instruction types (explained above) are used for many source-code patterns implementation, including loop breaks, \texttt{goto}, returning from functions, and terminating the program. Thus, the respective branch constraints are considered relevant as changing these branches direction may immediately affect the program control flow. In order to perform this check, we consequently disassemble all the instructions located in the branch scope, and examine them until the appropriate instruction is found or the instructions block ends.

\begin{table}[htbp]
\caption{Applying Strong Optimistic Strategy to Listing~\ref{lst:example}}
\begin{center}
\small
\begin{tabular}{c >{\columncolor[gray]{0.9}}c c >{\columncolor[gray]{0.9}}c c}
\toprule

    Iter&point&cs&c&\textbf{$\Pi_{sopt}$} \\
    \midrule
    0 & line 2 &
        \begin{tabular}{@{}l l@{}} 1 & main() \\
        2 & func() \end{tabular}
        & --- & $c_{2}$ \\
    \midrule
    1 & line 16 & \begin{tabular}{@{}l l@{}} 1 &
        main() \end{tabular} & buf[1] - buf[3] == 1 & $c_{2} \land c_{15}$ \\
    \midrule
    2 & line 16 & \begin{tabular}{@{}l l@{}} 1 &
        main() \end{tabular} & buf[0] == ’3’ & $c_{2} \land c_{15}$ \\
\bottomrule
\end{tabular}
\label{tbl:ex}
\end{center}
\end{table}

As an illustration, consider the example in Listing~\ref{lst:example}. Table~\ref{tbl:ex} demonstrates the algorithm execution on this example. The original sliced predicate looks as follows: $\Pi = (\texttt{buf[0]} = \texttt{'3'})\ \land\ (\texttt{buf[1]} - \texttt{buf[3]} = 1)\ \land\ (\texttt{buf[3]} = \texttt{'6'}\ \land\ \texttt{buf[0]} = \texttt{'5'})$. The target branch is located in line~2, and the initial $cs$ contains \texttt{func} and \texttt{main} call sites. On the first iteration, we analyze the branch in line~15. Since the call stack at that point contain only \texttt{main} function call site, we should update $point$ and $cs$ values. After that, we find out that reaching the $point$ value is control dependent on this branch, as its $src\_addr$ corresponds to line~15, and $dst\_addr$ is related to the end of \texttt{main} function (line~17). Hence, the constraint from line~15 is added to the predicate. On the second step, we consider the branch in line~13. The function frames are the same but the destination address of that branch is located before the $point$~-- in line~15. The target branch is not control dependent on the branch in line~13, and there are no CTI instructions in the latter's scope, thus constraint from line~13 is not added to the predicate. So, the final strong optimistic predicate has the following form: $\Pi_{sopt} = (\texttt{buf[3]} = \texttt{'6'}\ \land\ \texttt{buf[0]} = \texttt{'5'})\ \land\ (\texttt{buf[1]} - \texttt{buf[3]} = 1)$.

We suggest the following code to demonstrate the processing of control transfer instructions:
\begin{lstlisting}[language=C, basicstyle=\scriptsize\ttfamily, numbers=left, xleftmargin=2em, label=lst:cjson, caption={cJSON example with CTI processing.}, captionpos=b]
// cJSON.c:1100
// cJSON_ParseWithLengthOpts()
{
    ...
    item = cJSON_New_Item(&global_hooks);
    if (item == NULL) /* memory fail */
    {
        goto fail;
    }
    ...
    if (!parse_value(item,
    buffer_skip_whitespace(skip_utf8_bom(&buffer))))
    {
        ...
    }
    ...
fail: ...
}
...
// cJSON.c:1317
// parse_value()
{
    ...
    if (can_read(input_buffer, 4) &&
    (strncmp((constchar*)buffer_at_offset(input_buffer),
    "null", 4) == 0))
    {
        item->type = cJSON_NULL;
        input_buffer->offset += 4;
        return true;
    }
     /* false */
     ...
 }
\end{lstlisting}

\begin{table}[htbp]
\caption{Applying Strong Optimistic Strategy to Listing~\ref{lst:cjson}}
\begin{center}
\scriptsize
\begin{tabular}{c >{\columncolor[gray]{0.9}}c c >{\columncolor[gray]{0.9}}c c}
\toprule

    Iter&point&cs&c&\textbf{$\Pi_{sopt}$} \\
    \midrule
    0 & line 25 &
        \begin{tabular}{@{}l l@{}} 1 & cJSON\_ParseWithLengthOpts() \\
        2 & parse\_value() \end{tabular}
        & --- & $c_{25}$ \\
    \midrule
    1 & line 11 & \begin{tabular}{@{}l l@{}} 1 &
        cJSON\_ParseWithLengthOpts() \end{tabular} & $c_{6}$ & $c_{25} \land c_{6}$ \\
\bottomrule
\end{tabular}
\label{tbl:ex-cjson}
\end{center}
\end{table}

Suppose our goal is to make the constraint in lines~25--26 false (invert the corresponding branch). Note that a single if-statement in source code may contain several branches, divided by logical \texttt{\&\&} and \texttt{||} operators, as in this example. Each branch is expressed via its own \texttt{jcc} instruction, i.e. it is the individual branch in binary code. Suppose that the sliced predicate for inverting the branch turned out to be UNSAT, so, we try to apply the strong optimistic strategy. According to the algorithm, we add the target branch to the resulting predicate and save the call stack in that point ($point$ and $cs$ values are shown in Table~\ref{tbl:ex-cjson}). Then we start iterating over sliced constraints and at some moment get to the branch in line~6. As this branch is located in another function frame, we need to change the values of $point$ and $cs$. In this case call stacks differ only by \texttt{parse\_value} function, and its call site is in lines~11--12 (in function \texttt{cJSON\_ParseWithLengthOpts}). Thus, the current \texttt{point} is in line~11, and the analyzed branch scope is between lines~6 and 10, so, the $point$ is not control dependent on the analyzed branch. However, if we explore the scope instructions, we can see that there is \texttt{goto} command compiled as an unconditional jump to the label \texttt{fail} beyond the scope. As a result, the analyzed branch scope has an appropriate CTI inside, hence, the negation of constraint in line~6 should be added to the predicate.

The following Listing~\ref{lst:libcbor} shows another source code pattern that may lead to adding the constraint to the strong optimistic predicate, regardless of whether reaching the $point$ is control dependent on the corresponding branch:
\begin{lstlisting}[language=C, basicstyle=\scriptsize\ttfamily, numbers=left, xleftmargin=2em, caption={libcbor example with CTI processing.}, captionpos=b, label=lst:libcbor]
// cbor/internal/builder_callbacks.c:34
// _cbor_builder_append()
{
    assert(ctx->stack->top->subitems > 0);
    ...
    if (ctx->stack->top->subitems == 0) {
        ...
    }
    ...
 }
\end{lstlisting}

In this example, \texttt{assert} is actually a symbolic branch. When compiling it with clang without any optimizations, the following assembly is generated:
\begin{lstlisting}[basicstyle=\scriptsize\ttfamily, xleftmargin=2em, escapechar=!]
40618a: mov    rax,QWORD PTR [rbp-0x10]
40618e: mov    rax,QWORD PTR [rax+0x10]
406192: mov    rax,QWORD PTR [rax]
406195: cmp    QWORD PTR [rax+0x10],0x0
40619a: jbe    4061a5 <_cbor_builder_append+0xb5>
!\colorbox{light-gray}{4061a0: jmp    4061cd <\_cbor\_builder\_append+0xdd>}!
4061a5: movabs rdi,0x40eaa8
4061ac:
4061af: movabs rsi,0x40eac6
4061b6:
4061b9: mov    edx,0x22
4061be: movabs rcx,0x40eaf5
4061c5:
4061c8: call   4010a0 <__assert_fail@plt>
!\colorbox{light-gray}{4061cd: ...}!
\end{lstlisting}

Suppose we want to invert the branch in line~6 (e.g. make the condition to be false). Initially, $\Pi_{sopt}$ will contain the inverted constraint from line~6. Then at some moment we start to analyze the \texttt{assert} branch. The function frame is the same as in line~6, reaching the $point$ (that stands at line~6) is not control dependent on the branch. If we do not consider the CTI, we might miss this important branch and make the strong optimistic solution incorrect. But as the assembly has the exactly such form as in the listing, we can notice that an unconditional jump goes beyond the above branch destination address. Hence, \texttt{jmp} is an appropriate CTI that makes the above \texttt{jbe} to be added to the predicate.

\section{Implementation}
\label{implementation}

We implement strong optimistic solving in our dynamic symbolic execution tool Sydr~\cite{vishnyakov20}. Sydr consists of two processes: Concrete Executor, based on dynamic binary instrumentation tool DynamoRIO~\cite{bruening04} and Symbolic Executor, including dynamic symbolic execution framework Triton~\cite{saudel15}.

Concrete Executor runs the program, handles its instructions, and provides all the necessary information to the Symbolic Executor via shared memory events. On the concrete side, we use DynamoRIO to handle symbolic branches scopes to find control transfer instructions (CTI). When we meet a conditional branch, we get its source and destination addresses, and sequentially disassemble instructions between them checking their opcodes. For conditional and unconditional jumps we additionally check if their destination addresses go further the analyzed scope end. For each symbolic branch we fill the special field in the instruction event structure that signalizes whether this branch has one of the appropriate control transfer instructions inside its branch scope.

On the symbolic side, Sydr handles shared memory events. For the instruction event, it saves the CTI-field information and interprets symbolic instruction. If the instruction is \texttt{call} or \texttt{ret}, then Symbolic Executor updates the program call stack. Firstly, it removes inconsistent stack part, i.e. functions with \texttt{sp}-values greater than the current \texttt{sp}. Secondly, for the \texttt{call} instructions, it adds a new call stack entry with the function call site, current \texttt{sp}-value, instruction module address, and module id.

There are two main threads in Symbolic Executor: the first one uses Triton to build the path predicate, the second one inverts symbolic branches and communicates with Bitwuzla solver~\cite{niemetz20} to generate test cases. We implement optimistic and strong optimistic strategies in the second thread. We apply them while inverting symbolic branches in case the sliced path predicate turns out to be UNSAT. According to the flowchart in Figure~\ref{flowchart}, we ask the solver for the optimistic predicate satisfiability, and if it occurs to be SAT, we build the strong optimistic predicate and query the solver again.

\section{Evaluation}
\label{evaluation}

We evaluate strong optimistic solving efficiency and performance on a set of 64-bit Linux programs~\cite{sydr-benchmark} in the context of our dynamic symbolic execution tool Sydr~\cite{vishnyakov20}. The evaluation is performed in two steps. Firstly, we run benchmark and measure the analysis performance. Secondly, we collect the code coverage information to determine whether the strong optimistic strategy helps to discover new execution paths.

\begin{table*}[t]
\caption{Analysis Efficiency (Base)}
\begin{center}
\scriptsize
\begin{tabular}{l | @{\hspace{0.9\tabcolsep}} r >{\columncolor[gray]{0.9}}r r | r >{\columncolor[gray]{0.9}}r r | r >{\columncolor[gray]{0.9}}r r | r >{\columncolor[gray]{0.9}}r r}
\toprule
    \multirow{2}{*}{\textbf{Application}}&\multicolumn{3}{c}{\textbf{Default}}&\multicolumn{3}{c}{\textbf{Strong Optimistic Only}}&\multicolumn{3}{c}{\textbf{Optimistic Only}}&\multicolumn{3}{c}{\colorbox{lightgray}{\textbf{Optimistic + Strong Optimistic}}} \\
    &\textbf{Correct}&\textbf{Accuracy}&\textbf{Speed}&\textbf{Correct}&\textbf{Accuracy}&\textbf{Speed}&\textbf{Correct}&\textbf{Accuracy}&\textbf{Speed}&\textbf{Correct}&\textbf{Accuracy}&\textbf{Speed} \\

    bzip2recover&2101&100.0\%&135.55&2148&98.53\%&93.73&\textbf{2264}&100.0\%&141.8&\textbf{2264}&100.0\%&131.76  \\
    decompress&992&83.36\%&16.53&1550&65.73\%&25.83&\textbf{4063}&64.38\%&67.72&3091&60.79\%&51.52  \\
    faad&372&64.03\%&6.2&384&52.6\%&6.4&\textbf{409}&38.66\%&6.82&407&38.61\%&6.78  \\
    foo2lava&363&99.45\%&6.05&770&77.62\%&12.83&\textbf{779}&54.86\%&12.98&706&54.48\%&11.77  \\
    hdp&4172&68.38\%&69.53&7435&65.65\%&123.92&8091&66.83\%&134.85&\textbf{8379}&70.35\%&139.65  \\
    jasper&\textbf{19558}&98.33\%&325.97&16921&74.99\%&282.02&17961&61.59\%&299.35&16352&61.44\%&272.53  \\
    libxml2&1035&83.0\%&509.02&5708&85.91\%&2329.8&7896&89.32\%&3616.49&\textbf{7936}&89.77\%&3283.86  \\
    minigzip&3928&51.9\%&434.83&4110&49.93\%&441.94&\textbf{5105}&56.87\%&577.92&\textbf{5105}&56.87\%&539.26  \\
    muraster&2203&99.91\%&36.72&2282&92.65\%&38.03&\textbf{2465}&75.24\%&41.08&2461&74.73\%&41.02  \\
    pk2bm&190&99.48\%&670.59&194&73.76\%&554.29&273&53.42\%&963.53&\textbf{275}&53.82\%&750.0  \\
    pnmhistmap (ppm)&113&91.13\%&10.51&131&8.72\%&8.55&558&8.66\%&44.58&\textbf{571}&8.87\%&31.58  \\
    re2&471&100.0\%&7.85&745&23.68\%&12.42&1964&48.82\%&32.73&\textbf{2046}&52.29\%&34.1  \\
    readelf&648&46.99\%&10.8&1057&45.62\%&17.62&\textbf{2661}&59.08\%&44.35&2645&61.4\%&44.08  \\
    sqlite3&8412&99.98\%&3554.37&9500&99.62\%&2740.38&\textbf{10432}&35.31\%&3243.11&\textbf{10432}&35.31\%&2952.45  \\
    yices-smt2&4407&79.55\%&798.85&5170&76.77\%&937.16&6002&74.1\%&1087.98&\textbf{6006}&74.15\%&1085.42  \\
    yodl&1130&98.26\%&4520.0&2316&73.01\%&7720.0&3078&70.23\%&10260.0&\textbf{3257}&71.04\%&9771.0  \\
\bottomrule
\end{tabular}
\label{tbl:efficiency}
\end{center}
\end{table*}

We use a server with two AMD EPYC 7542 32-Core processors and 512GB RAM for Sydr benchmarking. For each target application we launch branch inversion for one hour. According to the updated Sydr architecture, symbolic execution is performed within the two main threads: the first one for the path predicate construction, and the second one for inverting symbolic branches and solving the queries. All encountered symbolic branches are inverted in the direct order. The SMT solver is working asynchronously with the job of the path predicate construction, and there are no time limits for the target program execution.

Tables~\ref{tbl:efficiency}~and~\ref{tbl:effiiciency-symptr} show the benchmarking results for symbolic execution with various strategies applied. Table~\ref{tbl:efficiency} contains results for the base symbolic execution mode with solving query timeout of 10 seconds, while Table~\ref{tbl:effiiciency-symptr} corresponds to the launch with symbolic pointer reasoning enabled (symptr mode) and solving query timeout of 30 seconds. In both cases we compared the following runs:
\begin{itemize}
    \item default symbolic execution without using any strategies,
    \item symbolic execution with optimistic solutions only,
    \item symbolic execution with strong optimistic solutions only,
    \item symbolic execution with both optimistic strategies applied.
\end{itemize}
For each configuration we measured the number of correct branches (column \textbf{Correct}), the analysis accuracy (column \textbf{Accuracy}), and the average number of correct branches discovered in one minute (column \textbf{Speed}).

To explore new execution paths, Sydr tries to invert symbolic branches encountered during the path predicate construction. Thus, the solver-generated input is called correct if it actually inverts the target symbolic branch, i.e. makes the execution go through the target branch in inverted direction compared to the original execution. The target program should reach this branch and pass through it in inverted direction on the correct test case. However, there may be both optimistic and strong optimistic solutions for target branch inversion. In this case we consider the branch to be correct when at least one of the optimistic predicates for its inversion has correct solution.

According to the flowchart in Figure~\ref{flowchart}, if the sliced predicate for inverting a branch turns out to be unsatisfiable, we have a chance to get both optimistic and strong optimistic solutions for this predicate. In this case, we consider them as one to count exactly the number of branches that have at least one satisfiable inversion predicate. And we check that at least one of them actually inverts the branch. If both queries are satisfiable, then the number of SAT queries is increased by two while we want to increase our counter by one. Similarly, if both solutions are correct, we get the correct solutions number instead of the correct branches number. Thus, we recalculate these numbers by subtraction of one in each case that leads to the wrong counting. After this operation we obtain the number of branches with satisfiable predicates and the number of correct branches (showed in column \textbf{Correct}). The analysis accuracy is calculated via dividing the number of correct branches by the number of SATs (column \textbf{Accuracy}). The symbolic execution speed is obtained as the result of dividing the number of correct branches by the symbolic execution time in minutes (column \textbf{Speed}).

First of all, Tables~\ref{tbl:efficiency}~and~\ref{tbl:effiiciency-symptr} show that the optimistic solving allows to discover more correct branches on all applications, except for \texttt{jasper}. Such results are obtained in both default and symptr symbolic execution modes. For \texttt{decompress}, \texttt{foo2lava}, \texttt{hdp}, \texttt{libxml2}, \texttt{pnmhistmap (ppm)}, \texttt{re2}, \texttt{readelf}, and \texttt{yodl} the number of correct branches increased in orders of magnitude. The decrease for \texttt{jasper} is caused by the additional time we spend for each UNSAT predicate to solve the optimistic predicate. As a consequence, we are not able to explore as much branches as in default mode. Moreover, some optimistic solutions may appear to be incorrect. Along with this, the number of correct branches discovered per time unit increased in all cases, except for \texttt{jasper} and \texttt{sqlite3}. Eventually, the optimistic strategy helps find much more correct branches with a significant increase in symbolic execution speed.

In Tables~\ref{tbl:efficiency}~and~\ref{tbl:effiiciency-symptr} we can also see that applying the strong optimistic strategy along with the optimistic one makes the results even better. On most applications the number of correct branches increased without significant speed loss, when compared to the optimistic-only configuration. We can notice that \texttt{libxml2} and \texttt{yodl} have the profit in both runs; \texttt{decompress}, \texttt{readelf}, and \texttt{sqlite3}~-- only in the symptr mode; \texttt{hdp}, \texttt{re2}, and \texttt{yices-smt2}~-- only in the base mode. However, for \texttt{faad}, \texttt{foo2lava}, \texttt{jasper}, and \texttt{muraster} this value became smaller in both base and symptr variants. It is a consequence of the method flowchart in Figure~\ref{flowchart}. In case the sliced predicate for inverting the branch is unsatisfiable, we frequently need to do much more additional work than querying the solver for just one branch satisfiability. Along with this, we construct the strong optimistic predicate and query the solver for its satisfiability. Hence, the more strong optimistics we solve, the more time for one branch we spend, and the less branches we are able to handle during one hour interval. So, if we have a limited time which is not enough for the complete application analysis (that is true for the programs with worse results) and the symbolic constraints are complex, the number of processed symbolic branches is very likely to decrease in comparison with the default symbolic execution mode.

\begin{table*}[t]
\caption{Analysis Efficiency (Symptr)}
\begin{center}
\scriptsize
\begin{tabular}{l | @{\hspace{0.9\tabcolsep}} r >{\columncolor[gray]{0.9}}r r | r >{\columncolor[gray]{0.9}}r r | r >{\columncolor[gray]{0.9}}r r | r >{\columncolor[gray]{0.9}}r r}
\toprule
    \multirow{2}{*}{\textbf{Application}}&\multicolumn{3}{c}{\textbf{Default}}&\multicolumn{3}{c}{\textbf{Strong Optimistic Only}}&\multicolumn{3}{c}{\textbf{Optimistic Only}}&\multicolumn{3}{c}{\colorbox{lightgray}{\textbf{Optimistic + Strong Optimistic}}} \\
    &\textbf{Correct}&\textbf{Accuracy}&\textbf{Speed}&\textbf{Correct}&\textbf{Accuracy}&\textbf{Speed}&\textbf{Correct}&\textbf{Accuracy}&\textbf{Speed}&\textbf{Correct}&\textbf{Accuracy}&\textbf{Speed} \\

    decompress&572&80.56\%&9.53&793&66.14\%&13.22&999&48.47\%&16.65&\textbf{1034}&50.27\%&17.23  \\
    hdp&1006&64.86\%&16.77&1311&64.33\%&21.85&\textbf{1432}&68.52\%&23.87&1419&68.39\%&23.65  \\
    jasper&\textbf{37227}&99.19\%&620.45&34845&88.43\%&580.75&35430&80.61\%&590.5&32972&80.62\%&549.53  \\
    libxml2&1184&91.22\%&215.93&6152&88.16\%&932.12&8674&88.13\%&1320.91&\textbf{8727}&88.67\%&1198.22  \\
    minigzip&577&59.73\%&9.62&604&60.04\%&10.07&\textbf{614}&60.02\%&10.23&\textbf{614}&60.2\%&10.23  \\
    re2&332&100.0\%&5.53&399&19.66\%&6.65&\textbf{1271}&49.73\%&21.18&1270&49.76\%&21.17  \\
    readelf&127&91.37\%&2.12&160&76.19\%&2.67&179&25.83\%&2.98&\textbf{207}&30.17\%&3.45  \\
    sqlite3&10338&99.98\%&1797.91&11502&99.47\%&1691.47&12462&39.44\%&1917.23&\textbf{12463}&39.45\%&1776.2  \\
    yices-smt2&2178&82.25\%&36.3&2326&74.74\%&38.77&\textbf{2734}&74.31\%&45.57&2561&74.38\%&42.68  \\
    yodl&1402&98.66\%&1682.4&2625&75.43\%&2812.5&3387&72.2\%&3628.93&\textbf{3566}&72.88\%&3626.44  \\
\bottomrule
\end{tabular}
\label{tbl:effiiciency-symptr}
\end{center}
\end{table*}

On the contrary, the configuration with strong optimistic solutions only leads to significant decrease in the number of correct branches and speed. When we stop employing optimistic solutions, we waste lots of potentially correct inputs that are generated on average several times faster than the strong optimistic ones. Besides, the queries for the strong optimistics are more likely to be unsatisfiable as they contain more than one constraint. Thus, we spend time and resources for solving more complex queries, and do not solve rather simple ones, what results in analysis efficiency decrease. So, the optimal approach is applying the strong optimistic strategy as a complement to the optimistic one.

Another observation is that the branch inversion accuracy decreases when applying any of the strategies, compared to the default mode, even when the absolute number of unique correct branches increases. The reason is that the corpus size enlarges significantly, especially with optimistic solutions used. Every unsatisfiable predicate leads to an attempt of generating a solution for the optimistic predicate that is much simpler than the original one, thus is more likely to be SAT. And in case it is, the strong optimistic solution may also be generated. But not all these solutions are really able to invert the target branch, so we get an impressive increase in the number of incorrect branches that were not encountered earlier. This directly affects the symbolic execution accuracy.

Overall, applying the combination of optimistic and strong optimistic strategies in most cases allows to discover more correct symbolic branches during analysis than in the optimistic-only configuration, and achieve the compatible performance.

Measuring the code coverage is the second evaluation step. We gather three corpuses containing solver-generated seeds provided by each configuration to estimate the difference in coverage between Base configuration, the one with optimistic solutions (Opt), and the one with both optimistic and strong optimistic solutions (Sopt). We use afl-showmap~\cite{fioraldi20} utility in QEMU-mode to get the total number of program edges achieved on files from each corpus. The results are shown in Table~\ref{tbl:code-coverage}. The columns \textbf{Base}, \textbf{Opt}, and \textbf{Sopt} show the number of program edges covered by the corpus for each configuration respectively. The column \textbf{Opt / Base} depicts the coverage growth provided by Opt (optimistic) configuration compared to Base. The column \textbf{Sopt / Opt} illustrates the coverage growth provided by Sopt (optimistic + strong optimistic) configuration compared to Opt.

As we can see, strong optimistic solutions provide additional coverage growth in comparison with the optimistic configuration on the majority of target applications: \texttt{cjpeg}, \texttt{decompress}, \texttt{faad}, \texttt{foo2lava}, \texttt{hdp}, \texttt{jasper}, \texttt{libxml2}, \texttt{minigzip}, \texttt{re2}, \texttt{readelf}. The growth range varies from insignificant (0.04\%~-- \texttt{cjpeg} and \texttt{foo2lava}) to quite large (7.28\%~-- \texttt{re2}). However, some programs do not gain the coverage growth from the strong optimistic solutions (\texttt{bzip2recover}, \texttt{muraster}, \texttt{pk2bm}, \texttt{pnmhistmap (ppm)}, \texttt{sqlite3}, \texttt{yices-smt2}, and \texttt{yodl}). There are two main reasons for this situation. Firstly, some predicates might have simple constraints that are not exposed to over- and underconstraint. In such cases, the optimistic solutions are enough to invert the corresponding branches. Secondly, some predicates might contain extremely complex constraints or be highly over-/underconstrained. In these situations strong optimistic strategy is still not sufficient to get the correct inputs. It is worth noting that most of these applications yet have nice benchmark results (in either base or symptr mode).

Concluding the performed evaluation, we can say that the strong optimistic strategy in combination with the optimistic strategy helps either increase the efficiency and speed of symbolic execution, or discover the new coverage. At the same time, there is no sense in using the strong optimistic strategy alone, as such variant leads to significant performance and accuracy loss.

\begin{table}[htbp]
\caption{Code Coverage}
\begin{center}
\scriptsize
\begin{tabular}{l @{\hspace{0.9\tabcolsep}} r r >{\columncolor[gray]{0.9}}r r >{\columncolor[gray]{0.9}}r}
\toprule
    \textbf{Application}&\textbf{Base}&\textbf{Opt}&\textbf{Opt / Base}&\textbf{Sopt}&\colorbox{lightgray}{\textbf{Sopt / Opt}} \\

    bzip2recover&281&282&+0.36\%&282&0\% \\
    cjpeg&2412&2427&+0.62\%&2428&+0.04\% \\
    decompress&4523&4535&+0.27\%&4679&+3.18\% \\
    faad&3772&4657&+23.46\%&4689&+0.69\% \\
    foo2lava&2396&2417&+0.88\%&2418&+0.04\% \\
    hdp&6979&7431&+6.48\%&7472&+0.55\% \\
    jasper&1556&1571&+0.96\%&1574&+0.19\% \\
    libxml2&8763&8914&+1.71\%&8928&+0.16\% \\
    minigzip&1675&1675&0\%&1696&+1.25\% \\
    muraster&3368&3394&+0.77\%&3394&0\% \\
    pk2bm&1572&1575&+0.19\%&1575&0\% \\
    pnmhistmap (ppm)&1561&1705&+9.22\%&1705&0\% \\
    re2&4101&4547&+10.88\%&4878&+7.28\% \\
    readelf&4086&4971&+21.66\%&5076&+2.11\% \\
    sqlite3&4674&4702&+0.6\%&4702&0\% \\
    yices-smt2&6091&6145&+0.89\%&6145&0\% \\
    yodl&3156&3174&+0.57\%&3174&0\% \\
\bottomrule
\end{tabular}
\label{tbl:code-coverage}
\end{center}
\end{table}

\section{Conclusion}
\label{conclusion}

We propose strong optimistic solving and implement it in our dynamic symbolic execution tool Sydr~\cite{vishnyakov20}. It allows to eliminate irrelevant constraints from the path predicate based on the information about symbolic branches nesting. Additionally, symbolic constraints, which may affect the program control flow, are not eliminated from the resulting predicate. The proposed method uses control dependency and call stack analysis to determine relevant branches.

The strong optimistic strategy allows to discover new execution paths by inverting more symbolic branches along the single execution trace. We evaluate the method on the set of real-world applications~\cite{sydr-benchmark} measuring the analysis efficiency (i.e. correctly inverted
branches number, accuracy, and speed) and the code coverage. According to the results, applying the strong optimistic strategy along with the optimistic strategy~\cite{yun18} leads to either discovering more correct symbolic branches at the same time without great performance loss, or increasing the explored code coverage after a complete program exploration. At the same time, there is no sense in using the strong optimistic strategy alone, as in this case significant performance and accuracy loss is detected. So, the optimal way is applying the combination of the solving strategies, depending on the corresponding queries satisfiability.

\printbibliography

@inproceedings{fioraldi20,
  title={{{AFL++}}: Combining Incremental Steps of Fuzzing Research},
  author={Fioraldi, Andrea and Maier, Dominik and Ei{\ss}feldt, Heiko and Heuse, Marc},
  booktitle={14th USENIX Workshop on Offensive Technologies (WOOT 20)},
  url={https://www.usenix.org/system/files/woot20-paper-fioraldi.pdf},
  year={2020}
}

@inproceedings{stephens16,
  title={Driller: Augmenting Fuzzing Through Selective Symbolic Execution},
  author={Stephens, Nick and Grosen, John and Salls, Christopher and Dutcher, Andrew and Wang, Ruoyu and Corbetta, Jacopo and Shoshitaishvili, Yan and Kruegel, Christopher and Vigna, Giovanni},
  booktitle={NDSS},
  volume={16},
  number={2016},
  pages={1--16},
  year={2016}
}

@inproceedings{poeplau21,
  title={{{SymQEMU}}: Compilation-based symbolic execution for binaries},
  author={Poeplau, Sebastian and Francillon, Aur{\'e}lien},
  booktitle={Proceedings of the 2021 Network and Distributed System Security Symposium},
  year={2021},
  doi = {10.14722/ndss.2021.23118},
}

@inproceedings{poeplau20,
  title={Symbolic execution with {{SymCC}}: Don't interpret, compile!},
  author={Poeplau, Sebastian and Francillon, Aur{\'e}lien},
  booktitle={29th USENIX Security Symposium (USENIX Security 20)},
  pages={181--198},
  url = {https://www.usenix.org/system/files/sec20-poeplau.pdf},
  year={2020}
}

@article{borzacchiello21,
  title={{{FUZZOLIC}}: mixing fuzzing and concolic execution},
  author={Borzacchiello, Luca and Coppa, Emilio and Demetrescu, Camil},
  journal={Computers \& Security},
  volume = {108},
  pages={102368},
  year={2021},
  publisher={Elsevier},
  doi = {10.1016/j.cose.2021.102368},
}

@article{baldoni18,
  author    = {Baldoni, Roberto and Coppa, Emilio and D'Elia, Daniele Cono and Demetrescu, Camil and Finocchi, Irene},
  title     = {A Survey of Symbolic Execution Techniques},
  journal   = {ACM Computing Surveys},
  volume    = {51},
  number = {3},
  articleno = {50},
  publisher = {ACM},
  doi = {10.1145/3182657},
  year = {2018}
}

@inproceedings{cha12,
 author = {Cha, Sang Kil and Avgerinos, Thanassis and Rebert, Alexandre and Brumley, David},
 title = {Unleashing {{Mayhem}} on Binary Code},
 booktitle = {Proceedings of the 2012 IEEE Symposium on Security and Privacy},
 series = {SP~'12},
 year = {2012},
 pages = {380--394},
 numpages = {15},
 doi = {10.1109/SP.2012.31},
 publisher = {IEEE Computer Society},
}

@inproceedings{cadar06,
  author = {Cadar, Cristian and Ganesh, Vijay and Pawlowski, Peter M. and Dill, David L. and Engler, Dawson R.},
  title = {{{EXE}}: Automatically Generating Inputs of Death},
  year = {2006},
  publisher = {ACM},
  doi = {10.1145/1180405.1180445},
  booktitle = {Proceedings of the 13th ACM Conference on Computer and Communications Security},
  pages = {322–335},
  series = {CCS '06},
}

@inproceedings{cadar08,
  title={{{KLEE}}: Unassisted and Automatic Generation of High-Coverage Tests
         for Complex Systems Programs},
  author={Cadar, Cristian and Dunbar, Daniel and Engler, Dawson R},
  booktitle={OSDI},
  volume={8},
  pages={209--224},
  url={https://static.usenix.org/events/osdi08/tech/full_papers/cadar/cadar.pdf},
  year={2008},
}

@inproceedings{yun18,
  title={{{QSYM}}: A Practical Concolic Execution Engine Tailored for Hybrid Fuzzing},
  author={Yun, Insu and Lee, Sangho and Xu, Meng and Jang, Yeongjin and Kim, Taesoo},
  booktitle={27th USENIX Security Symposium},
  pages={745--761},
  url={https://www.usenix.org/system/files/conference/usenixsecurity18/sec18-yun.pdf},
  year={2018},
}

@inproceedings{vishnyakov20,
  title = {Sydr: Cutting Edge Dynamic Symbolic Execution},
  author = {Vishnyakov, Alexey and Fedotov, Andrey and Kuts, Daniil and Novikov,
            Alexander and Parygina, Darya and Kobrin, Eli and Logunova, Vlada
            and Belecky, Pavel and Kurmangaleev, Shamil},
  booktitle = {2020 Ivannikov ISPRAS Open Conference (ISPRAS)},
  pages = {46--54},
  year = {2020},
  publisher = {IEEE},
  doi = {10.1109/ISPRAS51486.2020.00014},
}

@inproceedings{vishnyakov21,
  title = {Symbolic Security Predicates: Hunt Program Weaknesses},
  author = {Vishnyakov, Alexey and Logunova, Vlada and Kobrin, Eli and Kuts,
            Daniil and Parygina, Darya and Fedotov, Andrey},
  booktitle = {2021 Ivannikov ISPRAS Open Conference (ISPRAS)},
  pages = {76--85},
  year = {2021},
  publisher = {IEEE},
  doi = {10.1109/ISPRAS53967.2021.00016},
}

@inproceedings{kuts21,
  title={Towards Symbolic Pointers Reasoning in Dynamic Symbolic Execution},
  author={Kuts, Daniil},
  booktitle={2021 Ivannikov Memorial Workshop (IVMEM)},
  year={2021},
  organization={IEEE},
  pages={42--49},
  doi={10.1109/IVMEM53963.2021.00014},
}

@inproceedings{balakrishnan04,
  author={Balakrishnan, Gogul and Reps, Thomas},
  title={Analyzing Memory Accesses in x86 Executables},
  booktitle={International conference on compiler construction},
  year={2004},
  publisher={Springer Berlin Heidelberg},
  pages={5--23},
  doi={10.1007/978-3-540-24723-4_2},
}

@inproceedings{saudel15,
  author    = {Saudel, Florent and Salwan, Jonathan},
  title     = {{{Triton}}: A Dynamic Symbolic Execution Framework},
  booktitle = {Symposium sur la s{\'{e}}curit{\'{e}} des technologies de l'information
               et des communications},
  series    = {SSTIC},
  pages     = {31--54},
  url       = {https://triton.quarkslab.com/files/sstic2015_slide_en_saudel_salwan.pdf},
  year      = {2015}
}

@phdthesis{bruening04,
  title={Efficient, Transparent, and Comprehensive Runtime Code Manipulation},
  author={Bruening, Derek},
  year={2004},
  school={Massachusetts Institute of Technology, Department of Electrical
          Engineering and Computer Science},
  url = {https://www.burningcutlery.com/derek/docs/phd.pdf},
}

@article{niemetz20,
  author    = {Aina Niemetz and
               Mathias Preiner},
  title     = {Bitwuzla at the {SMT-COMP} 2020},
  journal   = {CoRR},
  volume    = {abs/2006.01621},
  year      = {2020},
  url       = {https://arxiv.org/abs/2006.01621},
  archivePrefix = {arXiv},
  eprint    = {2006.01621},
}

@misc{smt,
  author = {Clark Barrett and Pascal Fontaine and Cesare Tinelli},
  title = {{The SMT-LIB Standard: Version 2.6}},
  institution = {Department of Computer Science, The University of Iowa},
  year = 2017,
  url = {www.SMT-LIB.org},
}

@article{chipounov11,
  author = {Chipounov, Vitaly and Kuznetsov, Volodymyr and Candea, George},
  title = {{{S2E}}: A Platform for in-Vivo Multi-Path Analysis of Software Systems},
  year = {2011},
  publisher = {ACM},
  volume = {46},
  number = {3},
  doi = {10.1145/1961296.1950396},
  journal = {SIGPLAN Notices},
  pages = {265--278},
}

@inproceedings{kang11,
  title={{{DTA++}}: Dynamic Taint Analysis with Targeted Control-Flow Propagation},
  author={Kang, Min Gyung and McCamant, Stephen and Poosankam, Pongsin and Song, Dawn},
  booktitle={Proceedings of the Network and Distributed System Security Symposium},
  series = {NDSS '11},
  year={2011},
}

@article{king76,
 author = {King, James C.},
 title = {Symbolic Execution and Program Testing},
 journal = {Communications of the ACM},
 volume = {19},
 number = {7},
 year = {1976},
 pages = {385--394},
 numpages = {10},
 doi = {10.1145/360248.360252},
 publisher = {ACM},
}

@inproceedings{shoshitaishvili16,
 author = {Yan Shoshitaishvili and Ruoyu Wang and Christopher Salls and Nick Stephens and Mario Polino and Andrew Dutcher and John Grosen and Siji Feng and Christophe Hauser and Christopher Kruegel and Giovanni Vigna},
 booktitle = {2016 IEEE Symposium on Security and Privacy (SP)},
 title = {{{SOK}}: (State of) The Art of War: Offensive Techniques in Binary Analysis},
 year = {2016},
 pages = {138-157},
 doi = {10.1109/SP.2016.17},
}

@article{xianya21,
 title={{{LeanSym}}: Efficient Hybrid Fuzzing Through Conservative Constraint Debloating},
 author={Xianya Mi and Sanjay Rawat and Cristiano Giuffrida and Herbert Bos},
 journal={24th International Symposium on Research in Attacks, Intrusions and Defenses (RAID ’21)},
 year={2021},
 pages={62--77},
 doi={10.1145/3471621.3471852},
}

@book{howard06,
  title={The security development lifecycle},
  author={Howard, Michael and Lipner, Steve},
  volume={8},
  year={2006},
  publisher={Microsoft Press Redmond},
  url={http://msdn.microsoft.com/en-us/library/ms995349.aspx},
}

@mastersthesis{pak12,
  title={Hybrid Fuzz Testing: Discovering Software Bugs via Fuzzing and Symbolic
         Execution},
  school = {School of Computer Science Carnegie Mellon University},
  author={Pak, Brian S},
  year={2012},
}

@book{iso08,
  title={{{ISO/IEC}} 15408-3:2008: Information technology -- Security techniques --
         Evaluation criteria for IT security -- Part 3: Security assurance
         components},
  year={2008},
  publisher={ISO Geneva, Switzerland},
  url={https://www.iso.org/standard/46413.html},
}

@book{gost16,
  title={{{GOST R}} 56939-2016: Information protection. Secure software development.
         General requirements},
  publisher={National Standard of Russian Federation},
  year={2016},
  url={http://protect.gost.ru/document.aspx?control=7&id=203548},
}

@inproceedings{serebryany16,
  title={Continuous Fuzzing with {{libFuzzer}} and {{AddressSanitizer}}},
  author={Serebryany, Kosta},
  booktitle={2016 IEEE Cybersecurity Development (SecDev)},
  pages={157},
  year={2016},
  organization={IEEE},
  doi={10.1109/SecDev.2016.043},
}

@misc{sydr-benchmark,
  title = {Sydr benchmark},
  url = {https://github.com/ispras/sydr-benchmark}
}

\end{document}